# Superconductivity without Pairing?

Alan M. Kadin
Princeton Junction, NJ 08550
Email amkadin@alumni.princeton.edu

**Abstract**
Electron pairing is central to modern concepts of superconductivity, and provides a basis for the ubiquitous presence of h/2e in the theory and phenomenology. However, the lack of a consistent real-space picture of phonon-mediated pairing suggests that the pairing interpretation may be incomplete or even misleading. Alternatively, the present picture refers back to an old proposal by Fröhlich, whereby superconductivity is based on an induced electron charge density wave coupled to a one-dimensional static lattice distortion. In the new picture, dynamic three-dimensional charge-density waves couple to coherent standing-wave phonons, giving rise to a mobile superconducting ground state that does not pin on crystalline defects. Long-range order in the ground state is associated with localized electron orbitals in a correlated two-phase field, which provides for flux quantization in units of h/2e on the macroscopic scale larger than the coherence length, without the need for a macroscopic pair wave function. While much of the formalism for this picture may carry over from conventional BCS theory, this picture makes distinct predictions for observable structures on the microscopic and mesoscopic scales.

## I.  Background:  Cooper Pairs, Charge Density Waves, and Kohn Anomalies

It has been known since the early days of quantum theory that a system composed of many identical bosons, i.e., integer spin particles, may condense into a single ground state with a large-amplitude coherent wave function.  In contrast, fermions are half-integer-spin particles for which only one one particle (with a distinct spin quantum number) may fill each energy state up to the Fermi level.  A He-4 atom is composed of 6 fermions (two protons, two neutrons, and two electrons) with a total spin of zero, and hence effectively acts as a boson.  So it is no surprise that the ground state of He-4 is a coherent superfluid.

In contrast, the charge carriers in normal metals are single electrons with spin ½, so the ground state would seem to be an incoherent Fermi sea of independent electrons. However, most non-magnetic metals actually become superconductors below a critical temperature $T_c$.  If the electrons could effectively combine and form boson composites, then this would seem to provide a simple explanation for the superconducting ground state.  Furthermore, magnetic flux quantization with h/2e follows directly for a superfluid composed of particles of charge 2e [1], based only on a single-valued quantum phase, analogous to quantization of angular momentum in the Bohr atom.

Indeed, the derivation of the bound state of the Cooper pair in 1956 [2], based on the electron-phonon interaction, provided the foundation for the Bardeen-Cooper-Schrieffer (BCS) theory of superconductivity in 1957 [3].  This has been widely seen as settling the question of superconductivity, although the discovery of high-temperature superconductors such as the cuprates has opened up the field to interaction mechanisms other than phonons. Still, the central role of Cooper pairs has seldom been questioned.

However, for a phenomenon as fundamental as superconductivity, there is a surprising lack of a consistent real-space picture for the Cooper pair in the superconducting ground state. The most compelling picture that is sometimes presented [4] is shown in Fig. 1. Here, a point electron near the Fermi surface with momentum $mv_F = \hbar k_F$ moves through a lattice of positive ions. The ions distort from their ideal lattice sites towards the moving electron, but due to the large velocity of the electron and the relatively sluggish motion of the ions, an ionic distortion with a net positive charge occurs as a tail behind the moving electron. Another electron moving in the opposite direction along the same line creates a similar positive tail, and each electron is attracted to the tail of the other, creating the basis for a net attraction that overcomes the basic Coulomb repulsion of the two electrons. While this picture serves well to motivate the electron-phonon interaction, it is not an accurate representation of the interactions that go into the BCS theory. First, electrons are not point particles at the atomic level; they are waves with a wavelength comparable to the interatomic distance. Second, given how fast the electrons move apart, this picture is incompatible with the concept of a stable bound state. And third, in real superconductors, the density of electrons is such that many Cooper pairs (typically, of order millions) are present on the scale of the superconducting coherence length ($\xi_0 \sim 100$ nm) that these interactions take place. This picture gives no hint to how coherence among many overlapping electrons can arise.

In contrast, the orbital picture of Kadin [5], although derived from the BCS ground state, appears quite different. This consists of a spherical electron standing wave, with components $k = \pm k_F$, bound to dynamic lattice distortions with $k = 2k_F$ (Fig. 2), providing a stable bound state that may move freely, carrying lossless current. Section II shows how this picture may be extended [6] to provide the basis for a coherent many-electron ground state.

The BCS theory gives rise to an energy gap $2\Delta$ below $T_c$, for electron states at the Fermi surface. It is instructive to compare this gap to the energy gap that arises at the Brillouin zone boundary for insulators in the classic band theory of solids (Fig. 3) [7]. For the insulator, diffraction of an electron wave with wavevector $k = G/2$ from lattice planes with reciprocal lattice vector $G$ gives rise to a standing wave. This standing wave exhibits a charge distribution with the same $2\pi/G$ wavelength as that of the lattice. The standing wave with charge antinodes aligned with positive ions has reduced energy and forms the electron state just below the energy gap. In contrast, the standing wave aligned out of phase with the ions has increased energy, and forms the state just above the energy gap. The standing waves are localized states bound to the fixed lattice, so that they cannot move freely, and hence one has an insulator. For the superconductor, the origin of the energy gap is generally believed to be quite different, due to the binding energy of the Cooper pair. But note the apparent similarity of the electron standing wave in Fig. 3 with that in Fig. 2. In the new picture presented below, the origins are fundamentally similar.

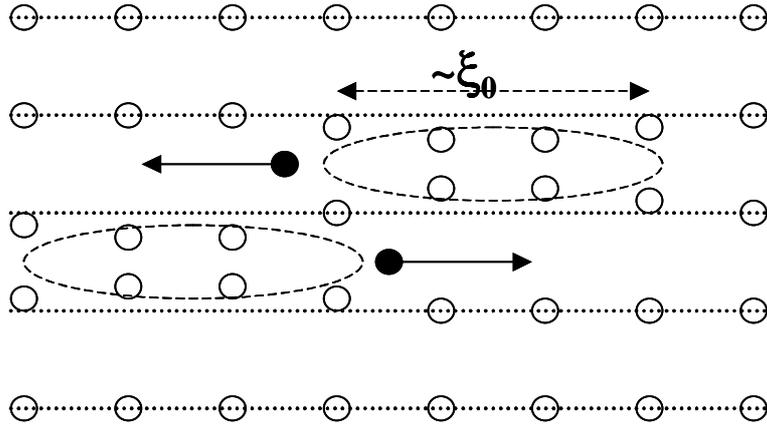

*Fig. 1. Conventional picture of electron-phonon interaction leading to formation of Cooper pair. Each rapidly moving electron leaves positively charged trail which can attract another electron moving in the opposite direction.*

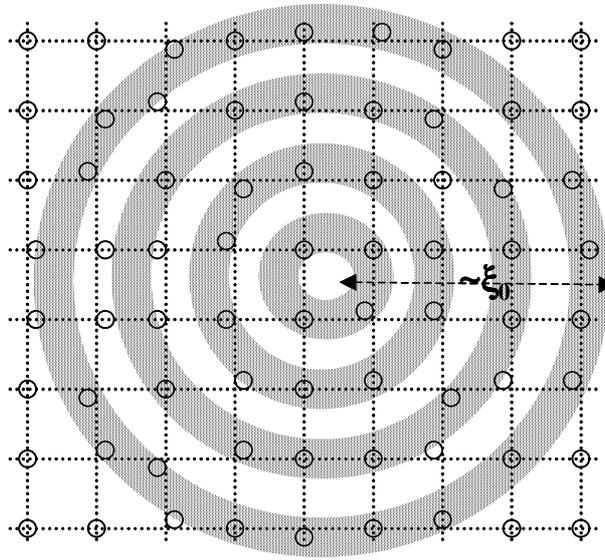

*Fig. 2. Localized electron orbital in superconducting state with components near Fermi surface, bound to standing-wave phonon.*

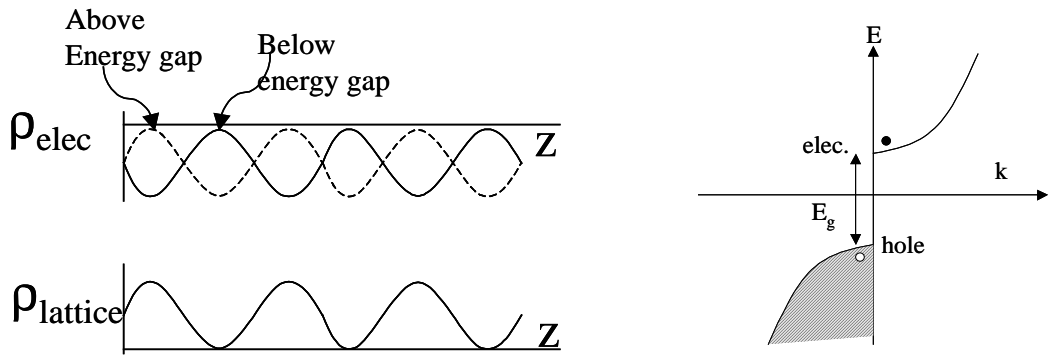

*Fig. 3. (Left) Electron and lattice charge distributions for Bloch waves at gap edge in insulators, charge density waves, and superconductors. (Right) Corresponding energy spectrum.*

While the pairing approach to superconductivity has been nearly universal, there have been other explanations proposed for superconductivity. While these were generally discarded, or kept as reference for exotic forms of superconductivity, one that has continued to be of interest relates to charge density waves [8,9]. In its essential form, a charge density wave (CDW) [10] is an electron standing wave that forms at the Fermi surface $k_F$, coupled to an induced one-dimensional static lattice distortion with $K = 2k_F$. (More generally $K = 2k_F-G$ for an appropriate reciprocal lattice vector G that brings K into the first Brillouin zone.) In effect, an induced lattice periodicity is formed that diffracts electron waves at the Fermi surface, thus creating electron standing waves with reduced energy much like those in the band insulator in Fig. 3. Such CDWs are found to occur below a critical temperature $T_c$ in some quasi-one-dimensional metals, and indeed they are insulators with an energy gap.

Fröhlich [8] proposed (even before the BCS theory) that a material with a CDW could form a superconductor rather than an insulator, if the lattice distortion and the electron standing wave that is bound to it were free to move with respect to the crystal lattice. However, in all known CDW materials, the lattice distortions pin on defects in the crystal lattice, and can move only with a large applied voltage, forming "sliding CDWs" that exhibit electrical dissipation. Hence real CDW materials are electrical insulators, and the Fröhlich mechanism for superconductivity was largely discarded.

Remarkably, the CDW phase transition and energy gap are described by identical equations to those of the BCS energy gap, based on the same electron-phonon interaction [10,11]. Furthermore, there have been several reports of coherent quantum phenomena in CDW materials, in particular evidence of flux periodicity of h/2e in sliding CDW states [12,13]. This 2e might appear to reflect pairing, except that the CDW ground state does not seem to involve pairs.

The lattice distortion coupled to a CDW can be viewed as a "frozen phonon", a limiting case of a phonon with $K=2k_F-G$ where the frequency has been driven to zero. More generally in metals not undergoing a CDW transition, much smaller changes in phonon frequency and lifetime are associated with states having $K=2k_F-G$, corresponding to scattering across the Fermi surface, and are known in the literature as "Kohn anomalies" [14]. Within the standard BCS picture, one does not expect Kohn anomalies to be associated with the superconducting state. However, a recent high-resolution neutron scattering study of the classic BCS superconductors Nb and Pb [15, 16] identified new Kohn anomalies in the phonon spectra, each with frequency such that $\hbar\omega = 2\Delta(0)$, the corresponding superconducting energy gap. This remarkable coincidence led the authors of that study to suggest the possibility of relations between superconductivity and charge density waves that go beyond conventional theory. While an alternative explanation based on conventional extensions of BCS was also suggested [17], a full understanding of these issues remains unresolved.

It is suggested here that the Fröhlich picture is indeed a good starting point for a more complete picture of superconductivity, and that the BCS theory really describes dynamic charge-density wave states in three dimensions, rather than bound electron pairs. Unlike a static one-dimensional CDW, the three-dimensional dynamic CDWs here correspond to high-frequency Kohn-anomaly phonons, such as those with $\hbar\omega = 2\Delta(0)$ identified in [15]. Further, these phonon states form a coherent dynamic "virtual phonon lattice", incommensurate with the crystal lattice, which binds a set of localized electron states (see Fig. 4). These dynamic CDWs do not pin on crystalline defects, and hence are free to move relative to the fixed lattice, carrying lossless supercurrent. So superconductivity and charge density waves are really aspects of the same physical phenomena.

But how can this one-electron band picture be compatible with the conventional superconducting phenomenology with paired charge carriers of 2e? It is proposed that this reflects structure on the mesoscopic level (on the scale of the superconducting coherence length ~ 100 nm), corresponding to long-range order with packing of localized electron states with spatially alternating quantum phase factors (Fig. 4), consistent with the Pauli exclusion principle for fermion states. This two-phase mesostructure leads on the macroscopic scale to the usual flux quantization with h/2e. This is described further in Section III.

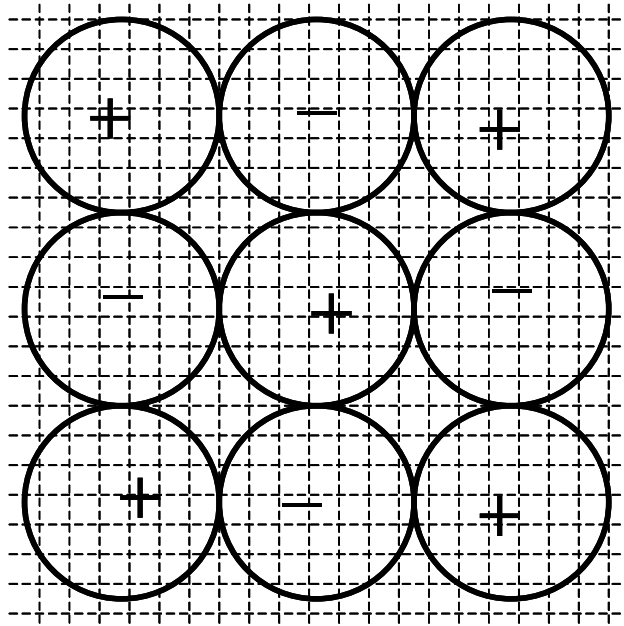

*Fig. 4. Conceptual picture of packing of localized electron states in superconducting ground state. Adjacent states of the same energy are mutually out of phase (phase shift of 180°). Electron states are tied to grid that represents "virtual phonon lattice", incommensurate with crystal lattice.*

Strictly speaking, this dynamic CDW picture of superconductivity corresponds to the electron-phonon interaction. However, this picture can be easily extended to describe superconductivity via dynamic spin density waves [18, 10], for example. Furthermore, the symmetry of the constituent localized orbitals can also differ, reflecting d-wave or other symmetries of the superconducting state [19]. In this way, this picture can provide a conceptual framework for a wide variety of conventional and exotic superconductors. This is discussed in Section IV.

Finally, is this picture merely an alternative interpretation of the BCS theory, or does this describe distinct physics? If the physics is distinct, what experiments can be used to verify this picture? It is suggested in Section V that attention be focused on two regimes: microscopic and mesoscopic. In the microscopic regime, the induced lattice periodicities, associated with scattering across the Fermi surface, should be observable by diffraction and scattering techniques. And in the mesoscopic regime, this picture predicts a crossover from a flux periodicity of h/2e to h/e for superconducting loops that are smaller than the coherence length.

## II. Microscopic Orbital Picture and Virtual Phonon Lattice

The orbital picture of Fig. 2 was derived [5] from the BCS equations, by Fourier transforming a ground-state distribution of electron waves within $\pm\Delta$ of a spherical Fermi surface. This directly yields a spherical S-wave orbital with nodes separated by $\sim \pi/k_F$ and an envelope that rolls off in a radius $\sim \hbar v_F/\Delta = \pi\xi_0$. Within the conventional pairing picture, this orbital represents the internal structure of a pair of bound electrons of opposite spin [4]. Such an orbital would be expected to form in the presence of a central attractive force between the two electrons. In fact, the direct interaction is repulsive and screened; a lattice modulation at $K = 2k_F$ (or equivalently, $2k_F$-G, i.e., a Kohn anomaly) is needed to diffract an outgoing wave at $k_F$ to an incoming wave at $-k_F$. But unlike the case of the band insulator, a static modulation with the appropriate properties is not present in the lattice.

Further, while an electron wave can scatter from a randomly fluctuating phonon, this would not maintain a bound state. Instead, what is needed here is a coherent oscillation whereby the electron waves adjust synchronously to maintain proper alignment with the nodes and antinodes in the lattice. Then the electron wave can be viewed as scattering coherently from the "virtual phonon", while the resulting electron standing wave (with charge modulation $2k_F$) induces and reinforces the phonon modulation at K. One may view each of the two electrons in the orbital as contributing coherently to the same phonon potential, hence they are effectively attracted to each other, thus comprising a pairing state. A large number of such identical Cooper pairs could then condense into a low-temperature Bose condensate state.

But this argument has a serious flaw. The modulated potential corresponding to two electrons is far too weak to account for the superconducting energy gap and critical temperature in real materials. Further, there may be millions of electrons on the scale of the coherence length $\xi_0$ that are part of the ground state. The only consistent picture is one in which all the electrons on the scale of $\xi_0$ are interacting with the same coherent set of lattice modulations. Given that real materials do not exhibit the spherical symmetry of Fig. 2, it is not immediately obvious how to obtain this coherence for a widely distributed set of orbitals. A one-dimensional set of diffracting planes, as suggested in ref. [6], is not consistent with the presence of an energy gap at the Fermi surface in all directions.

A close analogy with Bloch waves in a crystal resolves this question. The crystal has a set of lattice planes in three dimensions, and all of the Bloch waves reflect the periodicity in the lattice in their charge densities $\rho \sim |\Psi|^2$. This is true even though the various Bloch waves have different wave functions with different energies and frequencies, so that they are mutually orthogonal and incoherent. But their charge densities are all mutually coherent with the lattice. This creates an insulator with an energy gap at all parts of the Fermi surface.

Similarly, if one selects a set of equivalent locations on the Fermi surface in three dimensions, each corresponding to a Kohn-anomaly phonon with the same angular frequency $\omega_0$, then the set of planes creates a virtual (phonon) lattice. This virtual phonon lattice will have the same symmetry as the crystal lattice, but is incommensurate with respect to that lattice (see Fig. 5). This entire virtual phonon lattice also oscillates coherently at a single frequency $\omega_0$, with sites transforming into anti-sites and back again.

But for a nonspherical Fermi surface, the phonons that span the surface, with $K=2k_F-G$, will generally have significantly different frequencies $\omega(K)$ in different directions. Which value $\omega$ will be selected? A general rule may be that that value which provides for the largest reduction in band energy (the largest energy gap) will be automatically selected. However, the results of the recent experiments by Aynajian, which identified a new Kohn anomaly with $\hbar\omega = 2\Delta(0)$ for both Nb and Pb, suggest that this selection may turn out to be optimum, at least in some cases.

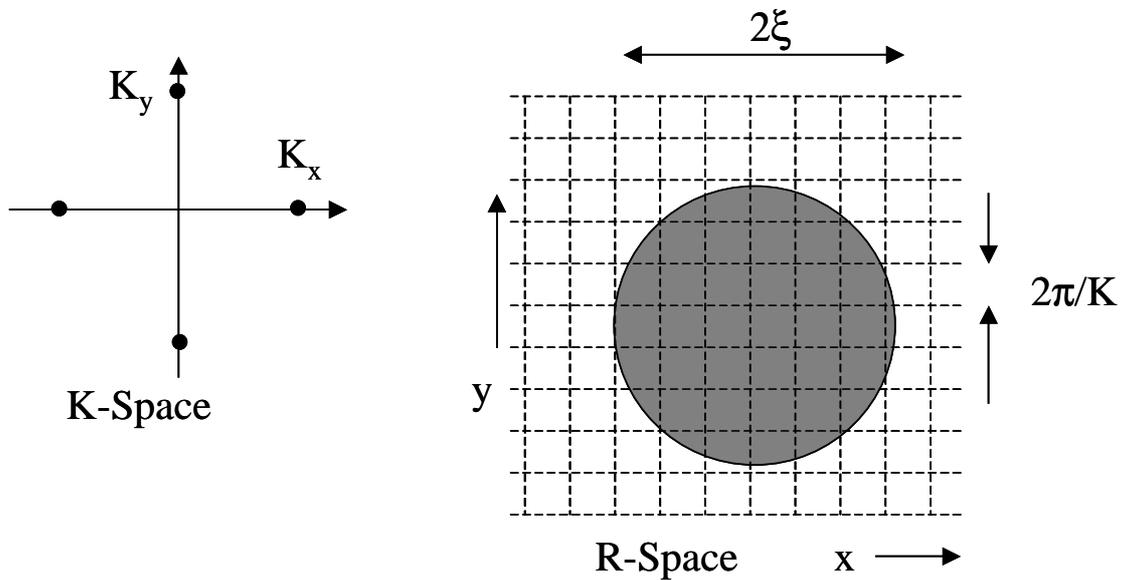

*Fig. 5. Conceptual picture of a virtual phonon lattice in K-space and R-space, with ground state orbital similar to that in Fig. 2.*

Like the case of the Bloch waves in the crystal, these various orbitals cover a range of energies. This is consistent with the Pauli principle for electron states that overlap in space. (But see Section III for the case of non-overlapping states of the same energy.)

Furthermore, the orbitals are not even all fixed with respect to the virtual phonon lattice. Recall that a Bloch wave with wavevector k near a Brillouin zone boundary at $k_Z$ has a net momentum $p = \hbar(k-k_Z)$. The same should be true for the Bloch waves in this virtual phonon lattice. While one could formally identify the two electrons on each orbital as a "pair", in fact they are no more paired than the Bloch waves in a crystal lattice. Furthermore, the fact that they exhibit different values of $\omega$ and k mean that they cannot correspond to a single energy level of a Bose condensate, so that the concept of coherent Cooper pairs loses its meaning.

It may be helpful to compare this microscopic picture of superconductivity to that for band insulators and CDW materials. For all of these cases, the localization scale is inversely proportional to the amplitude of the lattice modulation, which is proportional both to the reflection coefficient per plane, and to the energy gap. In a crystal lattice, of course, the modulation is quite large, and the typical energy gap is also large, of order 1 eV. In contrast, Nb and Pb have energy gaps ~ 2 meV, corresponding to a much weaker modulation. CDWs tend to lies between these limits. While the superconductor exhibits a coherence length and localization scale of order 100 nm, the corresponding length in CDWs and insulators would be expected to be proportionally smaller, perhaps only a couple of lattice spacings for the insulator.

Further, the localized states at the gap edge are tied to the relevant lattice. If this lattice can move in space, then the entire set of localized states can move with it, carrying electrical current which is lossless due to the energy gap. For a band insulator, of course, the carriers at the gap edge are tied to the crystal itself, so no current is possible. For the CDW case, a static CDW lattice pins easily on defects in the crystalline lattice, again preventing current flow. For the virtual phonon lattice, on the other hand, the high (~THz) oscillation frequency prevents pinning on static defects, and permits the entire virtual lattice and the attached electron orbitals to move in response to an electromagnetic field, thus producing true superconductivity.

As was pointed out above, the BCS gap equation applies equally well to superconductors and to charge density waves, so that the standard BCS formalism should continue to apply within this dynamic CDW picture. But the interpretation is quite different. Within the pairing picture, as one raises the temperature, pairs begin to be broken apart by thermal excitation, creating quasiparticles that weaken the ability of the remaining electrons to pair, thus reducing the energy gap. Within the dynamic CDW picture, the hole-like quasiparticle states are simply empty ground state orbitals, while the electron-like quasiparticle states are the "out-of-phase" Bloch states above the energy gap. There are no pairs to break; thermal excitation consists of removal of an electron from the ground state and placing it in an excited state, just as in the case of an insulator or semiconductor. These electron and hole excitations act like electrons and holes in semiconductors, and can produce resistance. But unlike the semiconductor case, the ground state here can carry dissipationless current. So the two-fluid model of superconductors should remain valid.

### III. Mesoscopic Orbital Packing and Macroscopic Phase Coherence

While the analysis above has shown that this dynamic CDW picture can account for zero resistance, the basis for macroscopic phase coherence is more subtle. This is in contrast to the case of the pairing picture, where the Bose condensation of quantum states of identical bound pairs directly provides synchronization of local quantum phase factors, each associated with charge 2e. This, in turn, provides a simple picture of magnetic flux quantization in units of h/2e. However, it is suggested here that this apparent simplicity is hiding detailed microstructure on the mesoscopic and microscopic scales, which is more complex but also more profound.

Fundamentally, the question is how a dense assemblage of electrons (or other fermions) can achieve macroscopic phase coherence, when the Pauli principle stipulates that two fermion wave functions which overlap in space cannot be coherent. And if they are not in contact, there is no apparent basis for them to be coherent. There is a simple way out of this conundrum, but one that does not seem to have been previously proposed. The Pauli principle permits two identical adjacent fermion wavefunctions to be in contact, as long as they are exactly 180° out of phase, with a node between them. In this way, when the two fermions are exchanged, the overall wavefunction is antisymmetric, i.e. changes sign. This antisymmetry upon particle exchange is well established as a characteristic of multi-fermion states. One can take this one step further, and tile space in a lattice with a basis of alternating phases (see Fig. 4). Now, each fermion is surrounded by N nearest neighbors with opposite phase. This is directly analogous to the crystal structure of an ionic solid, where each anion is surrounded by N cations and vice versa. Ionic crystals with one anion for each cation typically form in one of three cubic crystal structures: NaCl structure with 6 nearest neighbors; ZnS structure with 4 nearest neighbors, and CsCl structure with 8 nearest neighbors [7].

To the degree that the phase shift between adjacent sites is exactly 180°, such a two-phase mesostructure maintains a coarse-grainted long-range coherence of the quantum phase over an arbitrary distance. This does not involve pairing of fermions, and is not related to Bose-Einstein condensation. Nevertheless, it creates true macroscopic quantum coherence.

How does this fermion packing apply to the dynamic CDW picture of superconductors? One would expect that for each discrete Bloch wave near the energy gap, dense packing in the ground state would require this two-phase structure of orbitals on the scale of the coherence length, creating a second stage of organization on the mesoscopic scale, different from the first stage of the virtual phonon lattice on the microscopic scale. One would further expect that the packing lattices for each state would be similar to each other, but not necessarily aligned. In fact, the packing lattice is fixed relative to the virtual phonon lattice on the microscopic scale only for the top energy level, right at the energy gap. For other states below the gap, the properties of Bloch waves requires that all the orbitals of a given energy move together at some velocity relative to the virtual phonon lattice. Still, each level corresponds to a macroscopic coherent state with two reference phases, even though different levels will have different reference phases, and different energies will have relative reference phase differences that are changing in time.

Given the multiple different energy levels in the ground state, this is not a simple picture of macroscopic phase coherence with a unitary quantum phase $\phi$, or even a binary quantum phase $\phi$ and $\phi + \pi$. However, consider the situation if there is an electromagnetic field present, defined by a vector potential $\mathbf{A}(r)$. The phase factor for every electron orbital is described by the same canonical formula $\hbar \nabla \phi = m\mathbf{v_s} - e\mathbf{A}$. Let us first consider the case that $v_s=0$, which will turn out to correspond to a fixed microscopic virtual phonon lattice. Assume for simplicity a uniform $\mathbf{A}$ such that $\nabla\phi$ corresponds to a phase shift of 1° over a packing lattice distance of $2\xi$. Then the phases for the orbitals in the two-phase structure still alternate, but with an overall long-range phase shift of 1° per unit cell superimposed, as shown in Fig. 6. Now suppose further that the total phase shift for a chain of orbitals corresponds to 180°. One can take the two ends of this chain, and connect them while maintaining the phase gradient around the loop. (Note that this is 180°, not 360° as is required for a single coherent reference phase.) Using the identity that $\int \mathbf{A}\cdot d\mathbf{l} =\Phi$, the magnetic flux contained inside the loop, this leads directly to $\Phi = h/2e$. More generally, any integer multiple of 180° will also work, giving $\Phi=nh/2e$. Hence, one has magnetic flux quantization with $\Phi_0 = h/2e$, despite the fact that there are no Cooper pairs and no Bose condensation. This is due entirely to the two-phase mesostructure and the Pauli principle.

It is important to appreciate that each discrete energy level shows the same coherence with the same phase shift around the loop, although different energy levels may exhibit different orbital locations, and will in general exhibit net orbital motion relative to other energy levels. As long as each energy level interferes only with itself, everything is consistent.

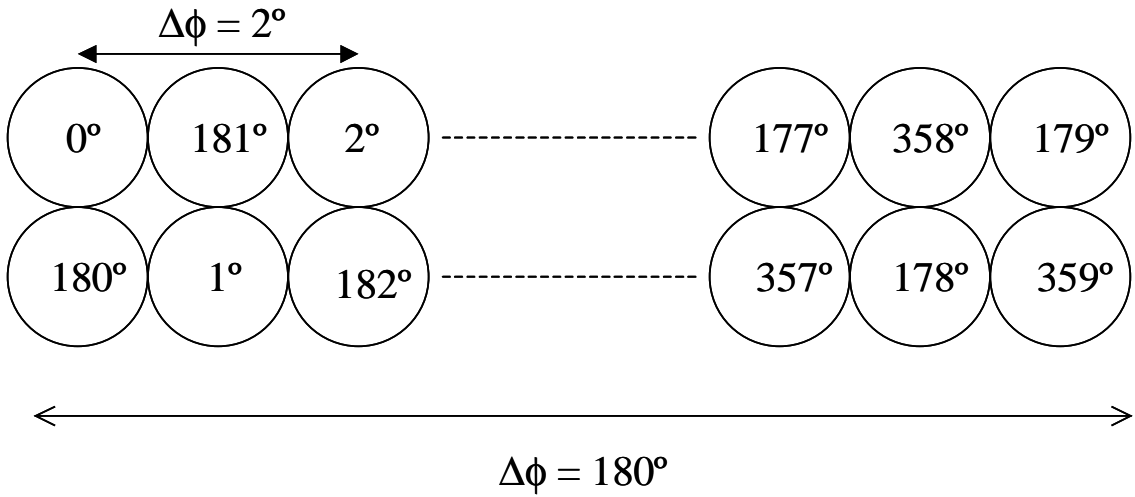

Fig. 6. *Conceptual picture of quantum phase gradient in fermion packing in magnetic field. The two ends can be reconnected, corresponding to a magnetic flux of h/2e.*

Now let us relax the constraint that $v_s=0$. This velocity corresponds to the net velocity of the virtual phonon lattice relative to the fixed crystal lattice (and the laboratory reference frame), and to shifting the entire Fermi surface relative to the fixed Brillouin zone. This is also the velocity term that goes into the electrical current, and hence can screen magnetic fields (see Fig. 7). Combined with flux quantization above, this gives rise to the usual Meissner effect and magnetic penetration depth $\lambda(T)$. This also provides the basis for the Ginzburg-Landau (GL) phenomenological equations, described by a macroscopic pseudo-wave function $\Psi = |\Psi|\exp(i\theta)$, where $|\Psi|$ is proportional to the amplitude of the coherent virtual phonon lattice vibrations, and $\theta$ is a reference phase such that $\nabla\theta$ is twice that for the single-electron reference phase. So $\Delta\theta$ around a loop is $2\pi n$, and $|\Psi| \propto \Delta$ as in the usual theory.

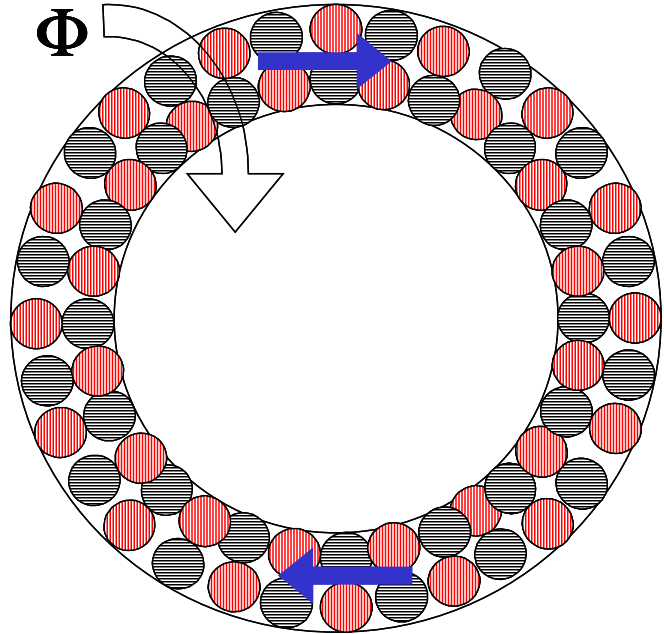

*Fig. 7. Conceptual picture of superconducting loop with screening current, corresponding to rotation of two-phase mesostructure.*

If the velocity $v_s$ is uniform across a superconducting sample, then the microscopic phonon lattice and the two-phase mesoscopic structure can move together as a rigid block. However, the picture becomes more complicated if $v_s$ is non-uniform. This occurs with screening of magnetic field, or when current is sent through a channel of non-uniform cross-section. In this case, parts of the structure must move at different velocities than other parts. This is consistent only if there are slip planes in the phonon lattice, that permit such relative motion while otherwise largely maintaining the local lattice structure. The coherence length $\xi$ represents the scale over which microscopic lattice coherence is necessary to achieve the full value of the energy gap. Gradients in $v_s$ over shorter distances, as occurs for example near a superconducting vortex, will weaken the energy gap (as indeed also follows from the GL equations). The macroscopic quantum phase coherence on larger scales does not require long-range coherence of the microscopic lattice, just that the two-phase packing is continuously maintained across the entire superconductor. Further, the two-phase packing need not exhibit rigid crystalline behavior at all; it may be more similar to an ionic liquid with short-range correlations, rather than to a solid ionic crystal.

### IV. Conventional and Exotic Superconductors

The model presented above is intended to describe a conventional metallic superconductor such as Pb or Nb, for which it is well accepted that phonons play a central role, and for which the energy gap is essentially isotropic. In these cases, s-wave orbitals in a three-dimensional phonon lattice at a single resonant frequency represent a consistent physical picture. But there are other kinds of superconductors as well, both observed and proposed. The present picture is easily generalized to include a variety of other, more exotic superconductors.

For example, consider the case of $MgB_2$, which is believed to have two discrete energy gaps, corresponding to separate parts of the Fermi surface [20]. As an extension of the present picture, one can consider two different phonon lattices, each based on different values of $\omega$ and K. Each section of the Fermi surface contributes orbitals in the ground state which are tied to the corresponding phonon lattice, and these two types of orbitals are only weakly coupled to each other. Thus, there is an energy gap around the entire Fermi surface, but the value of the energy gap differs significantly for different sections.

Another variant could involve orbitals with other than simple S-wave symmetry. For example, D-wave symmetry, believed to be present in most of the cuprate superconductors [19], has both anisotropy and in-plane angular nodes. The orbital packing in the present picture could reflect this symmetry, with phase shifts reflected both within the individual orbitals and between adjacent orbitals (see Fig. 8).

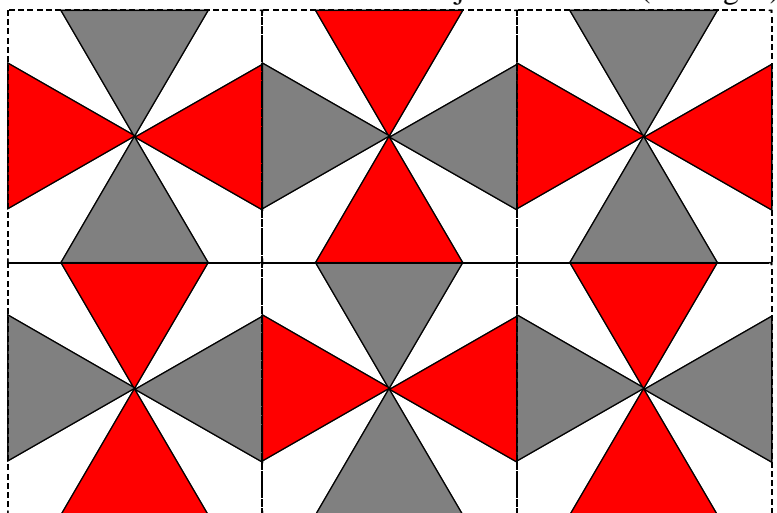

*Fig. 8. Picture of two-phase orbital packing for superconducting ground state with d-wave symmetry.*

Furthermore, this picture is not limited to an interaction based on conventional acoustic phonons. First, any other source of dynamic electrical polarizability coupling across the Fermi surface with $2k_F$-G could also contribute. One might envision, for example, coupling to optical phonons, surface plasmons, or excitons. A similar virtual lattice of dynamic polarization waves would yield a similar energy gap and superconducting ground state. Second, electrons have spin as well as charge. The analysis above implicitly assumed that electrons in both spin states of the orbitals behave identically. In contrast, if the ionic background exhibits substantial dynamic magnetic polarizability, i.e., spin waves corresponding to $2k_F$-G (Kohn anomalies), then the two spin states would be expected to diffract with opposite phases. This would lead to a dynamic spin-density wave (SDW), and to a "virtual magnon lattice" comparable to the virtual phonon lattice described earlier. The fundamental picture would be quite similar in both cases. Since high-temperature superconductors including both cuprates and iron pnictides seem to have magnetically active background states [21], this SDW variant may be quite relevant.

### V.    Experimental Validation of New Picture

As presented above, this dynamic CDW picture forms the basis of a more consistent and complete interpretation of the BCS and GL formalisms for superconductivity than the conventional pairing approach. Nevertheless, a more compelling argument would entail identification of specific experimental measurements that follow from this new picture, but would be unexpected within the pairing paradigm. While both low-temperature and high-temperature superconductors are of interest in this regard, it might be better to start with simple elemental superconductors such as Nb and Pb, where ideal samples are readily available and where it is well accepted (in both approaches) that phonons are responsible for superconductivity.

The high-resolution neutron scattering measurements of Aynajian et al. [15] present some compelling indirect evidence in support of this new picture. For both Pb and Nb, they have identified a new Kohn anomaly in the phonon spectrum, with frequency corresponding to the superconducting energy gap $2\Delta(0)$. These Kohn anomalies correspond to phonon wavelengths much longer than the lattice spacing in these metals, as summarized in Table I. Furthermore, Aynajian also found that in a Pb-Bi alloy where the energy gap shifted from pure Pb, the Kohn anomaly shifted correspondingly [16]. There are certainly other Kohn anomalies at much higher frequencies, but this apparent correlation suggests that the gap frequency is the relevant one for the dynamic CDW, perhaps associated with a resonant interaction that locks the frequencies together.

*Table I. Properties of Virtual Phonon Lattice*

|    | Structure | Crystal Lattice (Å) | Phonon Lattice (Å) | Frequency (GHz) |
|----|-----------|---------------------|--------------------|-----------------|
| Pb | fcc       | 5.0                 | 10                 | 730             |
| Nb | bcc       | 3.3                 | 29                 | 650             |

The question of Kohn anomalies in cuprates and other high-temperature superconductors is much less clear. This is combined with the uncertainty as to whether phonons are involved with superconductivity in these materials at all, or whether the focus should be on the magnon spectrum of some other set of excitations. One possibility is that some observations of a pseudogap in cuprates above $T_c$ [22] may actually reflect a Kohn anomaly in the appropriate excitation spectrum, rather than a direct indication of incipient superconducting order. In any case, further investigations into Kohn anomalies in cuprates and iron arsenides may be of interest.

As was discussed earlier, the dynamic CDW below $T_c$ corresponds to coherent standing-wave phonons with K and ω corresponding to the Kohn anomaly of interest. In a direct approach, one could carry out inelastic scattering with these values, and look for both emission and absorption peaks at low T. However, these peaks may be too weak to detect directly. Alternatively, it may be possible to use elastic scattering, i.e., diffraction, that covers a wide range of frequencies. For example, consider an x-ray detector in which an x-ray photon excites a core electron in much less than 1 ps. In this case, a spatial modulation that oscillates at 0.7 THz will appear essentially quasi-static, and will generate diffraction peaks in the usual way. As suggested in Fig. 9, one would expect to see a low-angle peak corresponding to K, as well as satellite peaks corresponding to G±K, where G is a peak in the crystal lattice. The intensities of both sets of peaks would be expected to follow the temperature dependence of the energy gap, going as $\Delta^2$, so a comparison above and below $T_c$ would help to identify even very small peaks.

The relative weakness of these coherent phonon diffraction peaks compared to the usual crystal diffraction peaks may be seen from the following argument. The amplitude of the energy gap for both superconductors and band insulators is proportional to the scattering potential, essentially the modulation depth. For the modulation depth in the crystal lattice, this leads to energy gaps of order 1 eV. In contrast, superconducting energy gaps may be up to a thousand times smaller. This suggests that the scattering amplitude for x-ray diffraction of coherent phonon peaks will similarly be a thousand times weaker than those from crystal peaks. In terms of diffracted intensity, this amounts to a very large factor of a million. This might initially seem to be an insurmountable difficulty, but high-intensity synchrotron x-ray sources have output intensities that may be six orders of magnitude higher than small laboratory x-ray sources. Still, it will be necessary to choose a regime where there is little or no background from crystal peaks not only of the crystal of interest, but also of weak impurity phases.

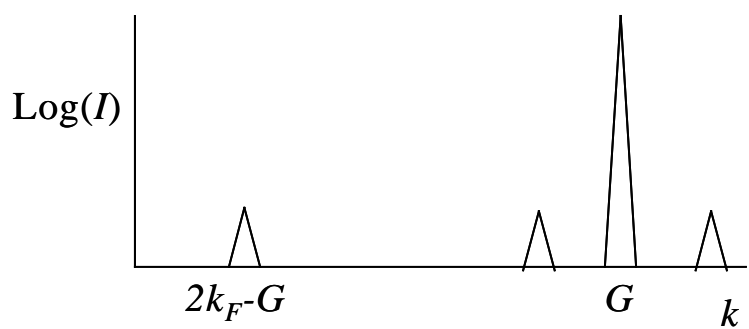

*Fig. 9. Conceptual diffraction scan of superconductor, showing weak low-k (low-angle) peak associated with virtual phonon lattice, as well as a crystal lattice peak with satellites from phonon modulation.*

In a material where the superconducting order is associated with spin waves rather than phonons, the virtual magnon lattice may be observed using polarized neutron scattering. Depending on the frequencies and detection methods, either inelastic scattering or diffraction may be applicable.

A completely different type of measurement can be undertaken to observe the mesoscopic orbital packing, on the scale of the coherence length. A diffraction probe on this scale would not be appropriate, since the packing may be more similar to a correlated liquid than to a crystal. On the other hand, the quantum intereference effect (modulation with magnetic flux) is the same for all energy levels, and gives rise to the usual periodicity in $\Phi_0$ for loops much larger than $\xi$. But this packing picture clearly suggests a crossover to a periodicity of $h/e = 2\Phi_0$ for smaller dimensions. It is hard to see how such a crossover would follow from the conventional understanding of the pairing picture. However, a recent theoretical analysis [23] has derived a similar crossover, based on the BCS formalism. It would be of great interest to carry out magnetization measurements on mesoscopic superconducting rings near and below $T_c$, where the coherence length is changing, to see if such a crossover is evident.

## VI.     Conclusions

A new picture of the superconducting state without Cooper pairs has been presented, based instead on a virtual phonon lattice coupled to a band of localized dynamic charge density wave orbitals. The phonon lattice corresponds to a Kohn anomaly at the frequency of the energy gap. The resulting ground state is much like that in a band insulator, except that the entire ground state can move freely with respect to the crystal lattice and collectively carry lossless current. The orbitals of each energy level in the ground state are packed with nearest neighbor states having alternating phases. The two-phase packing mesostructure permits long-range quantum phase coherence, which mimics the quantum coherence of a Bose condensate with charge 2e per particle. This picture can be directly extended to non-s-wave symmetry and other charged excitations. A further generalization can be made for magnetic excitations, where a virtual magnon lattice is coupled to a band of localized dynamic spin density wave orbitals. This generalized picture may apply to both low-$T_c$ and high-$T_c$ superconductors. While this picture has no Cooper pairs, the theory follows the usual BCS formalism, which also describes static density waves.

If the picture presented in the present paper is correct, this would bring about a dramatic revision in our understanding of the nature of the superconducting state and of long-range order in many-body fermion systems. In particular, it would suggest that macroscopic quantum coherence may actually be hiding a hierarchy of structures on different length scales. These structures are real and measurable, and should be accessible using scattering techniques and patterned nanoscale devices.

**Acknowledgment**
The author would like to thank Dr. Steven B. Kaplan of HYPRES, Inc. for his valuable insights into the nature of electrons and phonons in superconductors.